\documentclass[aps,prl,twocolumn,groupedaddress]{revtex4}

\usepackage[dvips]{graphicx,color} % Graphics and color
\usepackage{array,hhline,dcolumn} % Better table handling
\usepackage{rotating} % Rotate and sideways environments

\newcommand{\gtwid}{\mathrel{\raise.3ex\hbox{$>$\kern-.75em\lower1ex\hbox{$\sim$}}}}
\newcommand{\ltwid}{\mathrel{\raise.3ex\hbox{$<$\kern-.75em\lower1ex\hbox{$\sim$}}}}

\begin{document}
%\voffset=0.75in
%
%Below writes "DRAFT"
%\special{!userdict begin /bop-hook{gsave 200 30 translate
%65 rotate /Times-Roman findfont 216 scalefont setfont
%0 0 moveto 0.7 setgray (DRAFT) show grestore} def end}
% APS preprint designation
% \preprint{MiniBooNE-OSC}

%\title{Indication of Electron Antineutrino Appearance at the $\Delta m^2 \sim$ 1 $\mathrm{eV}^{2}$ Scale}
%\title{Indication of Electron Antineutrino Appearance Oscillations at the LSND Mass Scale}
%\title{Suggested $\bar \nu_\mu \rightarrow \bar \nu_e$ Oscillations at the 
%$\Delta m^2 \sim$ 1 $\mathrm{eV}^{2}$ Scale}
%\title{Further Evidence for $\bar \nu_\mu \rightarrow \bar \nu_e$ Oscillations at the 
%LSND Mass Scale}
%\title{Evidence for Electron Antineutrino Appearance Oscillations at the 
%LSND Mass Scale}
%\title{Excess Events in the Search for $\bar \nu_\mu \rightarrow \bar \nu_e$ Oscillations at the 
%$\Delta m^2 \sim$ 1 $\mathrm{eV}^{2}$ Scale}
\title{Corrections to the HARP-CDP Analysis of the LSND Neutrino Oscillation Backgrounds}

\author{
        G.~T.~Garvey, W.~C.~Louis, G.~B.~Mills, \& D.~H.~White
}
\smallskip
\smallskip
\affiliation{
Los Alamos National Laboratory; Los Alamos, NM 87545 \\
}

\date{\today}% It is always \today, today,

\begin{abstract}
Several mistakes have been found in recent papers that purport to reanalyze the backgrounds 
to the LSND neutrino oscillation signal. Once these mistakes are corrected, then it is 
determined that the background estimates in the papers are close to (if not lower than) the 
LSND background estimate. 
\end{abstract}


% Use showkeys class option if keyword
% display desired
\maketitle

The HARP-CDP group analyzed the pion production data taken by the HARP experiment at CERN with 
1.5 GeV/c protons incident on a Be target and performed a reanalysis \cite{harpcdp1,harpcdp2}
of the backgrounds to the 
LSND $\bar \nu_\mu \rightarrow \bar \nu_e$ 
oscillation signal \cite{lsnd_osc}. LSND observed a beam-on minus beam-off excess of
$117.9 \pm 22.4$ events. After subtracting a neutrino background of $30.0 \pm 6.0$ events, 
LSND determined the oscillation
signal to be $87.9 \pm 23.2$ events \cite{lsnd_osc}. The HARP-CDP group estimates a higher neutrino background
of $46.7 \pm 20.6$ events, which leads to an oscillation signal of $71.2 \pm 30.4$ events \cite{harpcdp2}. 
However, the HARP-CDP group made several errors 
in making their background estimate. The most egregious errors are discussed below.

HARP-CDP multiplies the intrinsic $\bar \nu_e$  background by a factor of 1.6, which is the 
ratio of ``Emulation''/``Best Estimate'' $\bar \nu_e$ in Table 15 \cite{harpcdp1}. 
However, this neglects the fact that 
HARP-CDP overestimates the decay at rest (DAR) fluxes and does not normalize to the $\nu_e$ flux. Thus, 
HARP-CDP instead should use a factor of 1.21, which is the ratio of ``Emulation''/``Best Estimate'' 
$\bar \nu_e/\bar \nu_\mu$ in Table 15. This increases the intrinsic $\bar \nu_e$ background by 4.1 events 
(from 19.5 to 23.6 total events) instead of by 11.7 events.  

In Table 15 of the first 
HARP-CDP paper \cite{harpcdp1}, the $\pi^+$ and $\pi^-$ decay in flight (DIF) fluxes are factors of 
3.3 and 2.5 higher in the ``Best Estimate'' than in the fluxes used by LSND \cite{lsnd_osc}. However, 
LSND has made high statistics measurements of $\nu_\mu$ and $\bar \nu_\mu$ scattering \cite{lsnd_cross},
and the HARP-CDP ``Best Estimate'' DIF fluxes are inconsistent with LSND measurements. For example, for 
$\bar \nu_\mu p \rightarrow \mu^+ n$ scattering, LSND observes $214 \pm 35$ events, which is consistent 
with the flux estimate and a factor of 3.3 times lower than the HARP-CDP flux estimate. For 
$\nu_\mu C \rightarrow \mu^- N_{gs}$ scattering, LSND observes $66.9 \pm 9.1$ events, which is consistent 
with the flux estimate and a factor of 2.5 times lower than the HARP-CDP flux estimate. Therefore, 
it is clear that HARP-CDP is overestimating the DIF flux and overestimating the number of DIF events 
observed in LSND by factors of 2.5-3.3. As the intrinsic $\bar \nu_e$ background all comes from $\pi^-$ DIF, 
this implies that the HARP-CDP intrinsic $\bar \nu_e$ background estimate should be reduced by a large factor 
(up to a factor of 3.3).  

In Table 17, the first HARP-CDP paper \cite{harpcdp1}
discusses the backgrounds from $\bar \nu_\mu p \rightarrow \mu^+ n$, 
where the $\mu^+$ is not observed if it is too low in energy ($T_\mu<3$ MeV). However, the $T_\mu<3$ 
MeV cut is not a hard cutoff. Rather, LSND still observes some muons down to 2 MeV or lower, especially 
because the energy lost by the recoil neutron is included. Also, LSND checked the background estimate by 
extrapolating the observed phototube (PMT) hit distribution down to zero. Therefore, the HARP-CDP background estimate 
of 13.8 events is overestimated, and HARP-CDP should use the LSND value of 10.5 events instead.

The second HARP-CDP paper \cite{harpcdp2}
estimates a background of 2.3 events from $\nu_e C \rightarrow e^- N_{gs}$ events, 
where the $N_{gs}$ beta decay mimics a 2.2 MeV $\gamma$ from neutron capture. 
However, this background is overestimated, partly 
because a 2.2 MeV positron produces more PMT hits than a 2.2 MeV $\gamma$. A 2.2 MeV $\gamma$ produces 
the same number of PMT hits as an $\sim 1-1.5$ MeV positron, including the energy from positron-electron
annihilation.           
In the LSND analysis, these $N_{gs}$ beta decays that mimic 2.2 MeV $\gamma$s are determined to be very 
small ($\sim 0.2$ events just to pass the minimal cuts). Indeed, LSND determined that the R distribution 
of $N_{gs}$ events looks indistinguishable from the R distribution of $N$ inclusive events without a beta.       

In summary, using the corrected estimate of the intrinsic $\bar \nu_e$ background of 23.6 events, 
the HARP-CDP excess should be 83.7 events, which agrees reasonably well with the LSND estimate of $87.9 \pm 23.2$ 
events. However, this ignores the problem with the HARP-CDP DIF fluxes, which overestimate the intrinsic 
$\bar \nu_e$ backround. For example, if the HARP-CDP intrinsic $\bar \nu_e$ flux is reduced by a factor of 
3.3 to make the DIF estimates agree with LSND data, then the HARP-CDP $\bar \nu_e$ background decreases to 
7.1 events and the excess increases to $100.2 \pm 23.2$ events.
%, corresponding to a $4.3 \sigma$ signal. 

\bibliography{dydak}% Produces the bibliography via BibTeX.

\end{document}